\documentclass[submission,copyright,creativecommons]{eptcs}
\providecommand{\event}{MARS 2018} 
\usepackage{underscore}           %
\usepackage{listings}
\usepackage{cadp-lnt}
\usepackage{enumerate}
\usepackage{courier}
\usepackage{multicol}
\usepackage{appendix}
\usepackage{graphicx}
\usepackage{caption}
\input{psfig}

\newlength{\magic}

\makeatletter
\def\Picture[#1]{\@ifnextchar[{\@lPicture[#1]}{\@Picture[#1]}}
\def\@Picture[#1]#2{\par\begin{center}\@lPicture[#1][14]{#2}\end{center}\par}
\def\@lPicture[#1][#2]#3{\mbox{\vspace*{1ex}\hspace*{0.8cm}\psfig{rheight=#1cm,rwidth=#2cm,bbllx=3.1714cm,bblly=0cm,bburx=0cm,bbury=\magic,figure=#3.ps}\hspace*{-0.8cm}\vspace*{1ex}}}
\makeatother

\makeatletter
\def\Figure{\@ifnextchar[{\@lFigure}{\@Figure}}
\def\@Figure#1{\begin{center} \input{#1} \end{center}}
\def\@lFigure[#1]#2{\hspace*{#1}{\parbox{\textwidth}{\@Figure{#2}}}}
\makeatother

\usepackage{subcaption}
\usepackage[symbol]{footmisc}

\def\mylabel#1#2{\@bsphack\if@filesw {\let\thepage\relax
   \def\protect{\noexpand\noexpand\noexpand}%
   \edef\@tempa{\write\@auxout{\string
      \newlabel{#1}{{#2}{\thepage}{}{figure.1.1}{}}}}%
   \expandafter}\@tempa
   \if@nobreak \ifvmode\nobreak\fi\fi\fi\@esphack}
\makeatother

\title{A Formal TLS Handshake Model in LNT}

\author{
Josip Bozic
\institute{Graz University of Technology}
\institute{Institute of Software Technology\\
Graz, Austria}
\email{jbozic@ist.tugraz.at}
\and
Lina Marsso
\institute{Univ. Grenoble Alpes, Inria, CNRS, Grenoble INP\footnote{Institute of Engineering Univ. Grenoble Alpes}, LIG\\ 38000 Grenoble, France}
\email{lina.marsso@inria.fr}
\and
Radu Mateescu
\institute{Univ. Grenoble Alpes, Inria, CNRS, Grenoble INP\footnotemark[\value{footnote}], LIG\\ 38000 Grenoble, France}
\email{radu.mateescu@inria.fr}
\and
Franz Wotawa
\institute{Graz University of Technology}
\institute{Institute of Software Technology\\
Graz, Austria}
\email{wotawa@ist.tugraz.at}
}

\def\titlerunning{A Formal TLS Handshake Model in LNT}
\def\authorrunning{J. Bozic, L. Marsso, R. Mateescu \& F. Wotawa}
\begin{document}
\sloppy
\maketitle

\begin{abstract}
Testing of network services represents one of the biggest challenges in cyber security. Because new vulnerabilities are detected on a regular basis, more research is needed. These faults have their roots in the software development cycle or because of intrinsic leaks in the system specification. Conformance testing checks whether a system behaves according to its specification. Here model-based testing provides several methods for automated detection of shortcomings. The formal specification of a system behavior represents the starting point of the testing process. In this paper, a widely used cryptographic protocol is specified and tested for conformance with a test execution framework. The first empirical results are presented and discussed.
\end{abstract}

\section{Introduction}

Security services are frequently used in fields like online banking, e-government and online shops. With increased availability of such services the number of security risks rises both for users and providers alike. In order to ensure a secure communication between peers in terms of authenticity, privacy and data integrity, cryptographic protocols are applied to regulate the data transfer. These protocols provide a standardized set of rules and methods for the interaction between peers. 

Transport Layer Security (TLS) \cite{dierksrescorla} is a widely used security protocol. TLS is the successor of Secure Sockets Layer (SSL) \cite{weaver}. Both protocols encompass a set of rules for the communication between client and server and rely on public-key cryptography in order to ensure integrity of exchanged data. 

However, despite multiple prevention measurements several vulnerabilities, like Heartbleed \cite{durmericetall} and DROWN \cite{aviramandall}, among others, have been discovered recently. This leads to the conclusion that more effort has to be invested for testing the implementations of such security protocols.

For this sake, many approaches have been introduced over the years. Some of them come from the area of model-based testing. Methods like fuzzing \cite{ruiterpoll} encompass methods and principles and help to detect further leaks in software. Other techniques rely on evolutionary algorithms \cite{Brubakerandall} or adaptations of artificial intelligence to concrete types of a system under test (SUT). 

On the other hand, non-functional testing, i.e, testing the way that the system operates, like conformance testing \cite{tretmans}, is applied to check whether a system corresponds to its specification.

In this paper, we present a formal model for the draft TLS 1.3 handshake \cite{draft-ietf-tls-tls13-21}, defined according to the TLS standard.
According to our knowledge, this is the first formal model of the draft TLS 1.3 handshake.
Then, a test execution framework tests a TLS implementation in an automated manner and checks whether the execution is conform to the behaviour specified by the formal model. Finally, a verdict is given about the correctness of the tested TLS implementations in terms of the obtained test results.

The remainder of the paper is organized as follows. Section~\ref{sec:handshakeTLS} gives an overview of the TLS 1.3 handshake. Section~\ref{sect:formalmodel} describes the challenges and choices for devising our formal model. Section~\ref{sec:validation} describes our validation approach and discusses the results.  Section~\ref{sec:rel} gives an overview about related literature. Finally, Section~\ref{sec:conc} gives some concluding remarks and future work directions.

\section{Transport Layer Security Handshake Protocol}
\label{sec:handshakeTLS}

The handshake protocol enables a TLS client and server to establish a secure, authenticated communication link.
The TLS handshake consists of the four steps: (i) consent on the version of the protocol to use and choose cryptographic algorithms; (ii) exchange and validate certificates to authenticate each other; (iii) generate a shared secret key; and (iv) abort the handshake with an alert.

\subsection{Main TLS 1.3 handshake messages}
The handshake itself consists of different message types, so-called TLS messages, and corresponding parameters that are part of these messages. Every such parameter comprehends specific values, where some of them are assigned dynamically during the handshake procedure.
The main messages exchanged during the TLS handshake steps are:
\begin{enumerate}[i.]
  \item The client and the server agree upon the version of the protocol and cryptographic algorithms to use by exchanging the {\tt client hello}, {\tt hello retry request}, {\tt server hello}, and {\tt encrypted extensions} messages. 
  \begin{itemize}
    \item The {\tt client hello} is always the first message, and a client should resend a {\tt client hello} message only if the server responded to it by a {\tt hello retry} request message.
    It contains the client's cryptographic information; the supported version of protocol, the pre-shared keys, the list of symmetric cipher options, and the extended functionalities. 
    \item The {\tt server hello} is the response (message) from the server to the {\tt client hello} message if the server was able to negotiate an acceptable set of handshake parameters based on the {\tt client hello}.
    It contains the cryptographic information negotiated: the protocol version, the list of symmetric cipher, and the server extensions.
    \item The {\tt hello retry request} is the response (message) from the server to the {\tt client hello} message if the server was not able to find an acceptable set of parameters. 
    It contains the same cryptographic information as the {\tt serverhello} message.
    \item The {\tt encrypted extensions} message is sent by the server immediately after the {\tt server hello} message.
    This is the first message that is encrypted using keys derived from the server\_handshake\_traffic\_secret.
    It contains extensions that can be protected.
  \end{itemize}
  \item An authentication with a certificate between the server and the client can be requested by exchanging the {\tt certificate request}, {\tt certificate}, and {\tt certificate verify} messages. 
  \begin{itemize}
  \item The {\tt certificate request} message must be sent by the server directly after the {\tt encrypted extensions} message if the server requests a certificate.
  It contains an identification of the certificate request and a set of extensions describing the parameters of the certificate. 
  \item The {\tt certificate} message is sent by the server and by the client.
  It contains their respective certificate to be used for authentication, and any other supporting certificates.
  \item The {\tt certificate verify} message is directly sent by the server and by the client after the respective certificate message.
  It contains a signature using the private key corresponding to the public key in the {\tt certificate} message. 
  \end{itemize}
  \item The server and the client generate and share a secret key by exchanging the {\tt finished} message. 
  The {\tt finished} message is sent by the server, then by the client.
  It contains a message authentication code (MAC).
  \item The client or the server can abort the handshake, and close the connection if at any step of the handshake, a failure happened.
  To do so, they should exchange an {\tt alert} message.
  The \emph{alert} message contains the description of the alert.
\end{enumerate}

\subsection{Handshake TLS 1.3 interactions}
\label{subsec:interactions}
The TLS messages make up the interaction between a client and a server and exchange values. The most important feature is the negotiation of cryptographic parameters that ensures privacy and integrity of exchanged information. During the procedure, client and server agree on used protocol version, exchange random values and select cryptographic algorithms for encryption and decryption of transferred data. Both peers exchange keys and certificates and after the handshake is finished, they can start to encrypt and exchange application data.

\begin{figure}
\begin{subfigure}{0.5\textwidth}
\center
\includegraphics[scale=0.4]{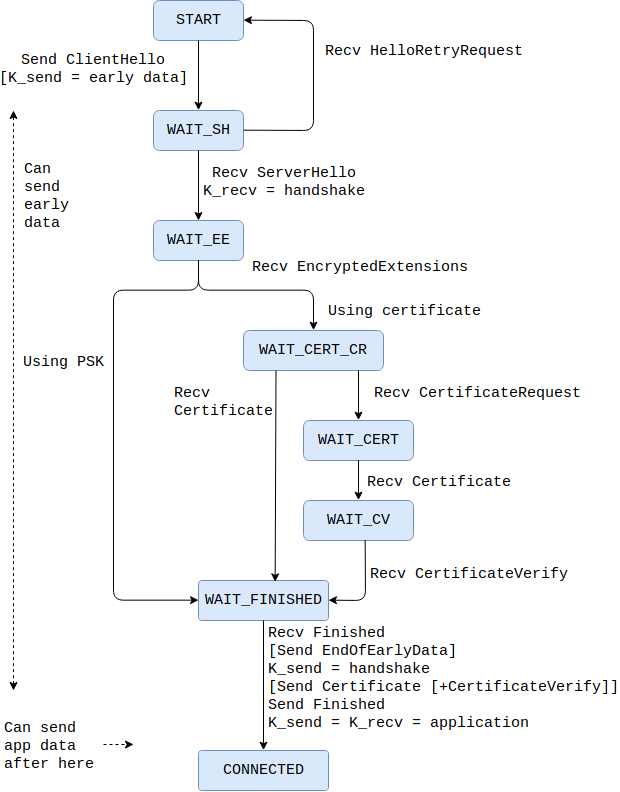} 
\caption{client}
\label{fig:statemachinesC}
\end{subfigure}
\begin{subfigure}{0.5\textwidth}
\center
\includegraphics[scale=0.4]{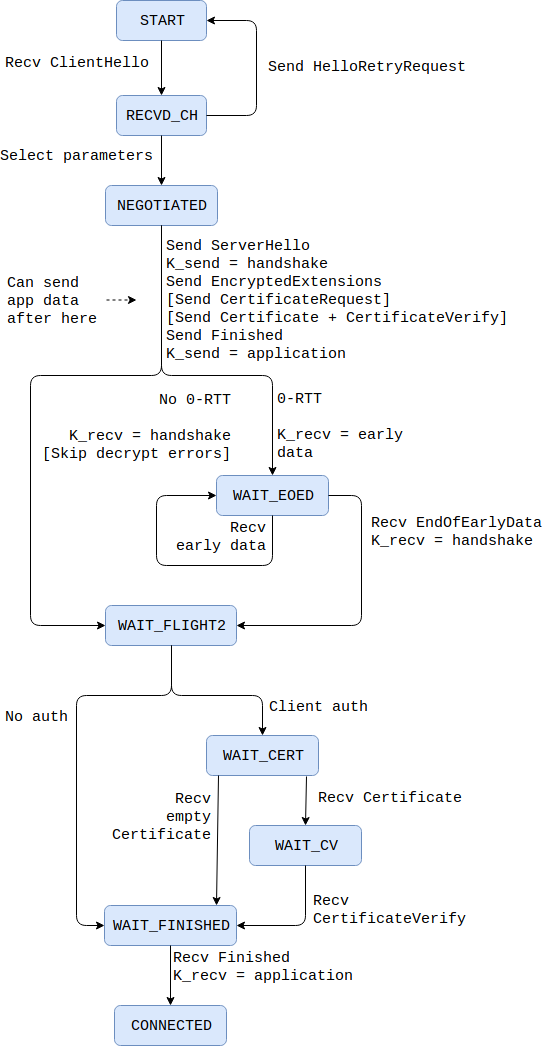} 
\caption{server}
\label{fig:statemachinesS}
\end{subfigure}
 
\caption{TLS handshake  client (a) and server (b) state machine (taken from \cite{draft-ietf-tls-tls13-21})}
\label{fig:statemachines}
\end{figure}

A summary of the interactions and the message exchanges from the client side and the server side is respectively depicted in the state machines~\cite{draft-ietf-tls-tls13-21} shown on Figure \ref{fig:statemachinesC} and Figure \ref{fig:statemachinesS}.
The state machines are transition based, the state names are for sake of readability only, and conditional actions are represented in brackets ([]).
Note that these state machines do not represent the client's and server's {\tt Alert} message interactions.
The server and the client have to abort the handshake with an {\tt Alert} message if one of the TLS 1.3 requirements textually described in the draft TLS 1.3 handshake \cite{draft-ietf-tls-tls13-21} is not respected.
As an example, we describe here three requirements taken from the draft TLS 1.3 handshake \cite{draft-ietf-tls-tls13-21}:
\begin{itemize}
    \item The handshake messages should be sent in one of the orders represented in a path of the state machines given in Figure \ref{fig:statemachines}.
    If a message is sent in the wrong order, the handshake connection will be aborted with an ``unexpected_message'' alert.
    \item The TLS 1.3 handshake refuses renegotiation without a {\tt hello retry request} message, thus the {\tt client hello} message can only be exchanged in the beginning of the protocol or after receiving a {\tt hello retry request} message.
    If renegotiation takes place, the handshake is aborted with an ``unexpected_message'' alert.
    \item When the client receives a {\tt hello retry request} message, the client should check that cryptographic information contained in the {\tt hello retry request} is different from the information in the initial {\tt client hello} message.
    If not, the handshake is aborted with an ``illegal_parameter'' alert.
\end{itemize}

In the sequel, we will not consider the internal processing of the TLS handshake but we will focus instead on the TLS messages, deaving the information exchange to be handled by the execution framework and the SUT.
Thus, our formal model represents the behaviour of the handshake, a critical part of the TLS.

\section{Formal Model of the TLS Handshake in LNT}
\label{sect:formalmodel}
Taking the draft specification of the Transport Layer security (TLS) \cite{draft-ietf-tls-tls13-21} protocol Version 1.3 as starting point, our main contribution in this paper is the formalisation of the Handshake protocol of TLS in the LNT language \cite{garavellangserwe17,champelovier2017reference}.
In this section, we give a brief description of our model,
which encompasses the handshake messages and illustrate then the handshake interactions. We discuss the challenges to specify the TLS handshake in LNT with concrete examples.
The full LNT model is given in Appendix \ref{ap:model}.

\subsection{Handshake messages}
\label{subsec:datastructure}
The handshake messages and their encryption information are defined as types.
Our model contains 43 types, with simple types as enumerations, lists, or more sophisticated ones, such as ``union like'' types.
\begin{lstlisting} [language=LNT]
type ClientHello is
  ClientHello (legacy_version: ProtocolVersion, random: Random32, legacy_session_id: SessionId, cipher_suite: Ciphers, legacy_compression_methods: CompressionMethods, extensions: Extensions)
end type
\end{lstlisting}
The main challenge was the definition of ``abstract'' types, regrouping several subtypes.
One of the encryption parameters of the {\tt client hello} message, the \emph{Extensions} is for instance an ``abstract'' type. 
The \emph{Extensions} is a list of \emph{Extension} types, and an \emph{Extension} is a tuple: an extension type and an extension data.
\begin{multicols}{2}
\begin{lstlisting} [language=LNT]
type Extensions is
  list of Extension
with "cons", "remove"
end type
\end{lstlisting}
\columnbreak
\begin{lstlisting} [language=LNT]
type Extension is
  Extension (extension_type: ExtensionType, extension_data: ExtensionData)
end type
\end{lstlisting}
\end{multicols}
LNT provides some predefined functions, which simplify the modeling task by avoiding the definition of classical useful functions.
For instance, the following LNT definition of the {\tt client hello} message described in Section \ref{sec:handshakeTLS}, uses three predefined functions to support notation x.f and compare values of the ClientHello type (``cons'' and ``remove'').
The extension type is defined by an enumeration of 21 extension types. The extension data is a type regrouping several constructors, each of them having its own parameters and corresponding to an extension type.
We implemented 9 of the 21 respective extension data constructors in our model.
\begin{multicols}{2}
\begin{lstlisting} [language=LNT]
type ExtensionType is
  signature_algorithms,                
  supported_versions,                  
  cookie,                                  
  ...
end type
\end{lstlisting}
\columnbreak
\begin{lstlisting} [language=LNT]
type ExtensionData is
  Cookie (c: Cookie),
  CertificateType (ct: CertificateType),
  SupportedVersions (sv: supportedVersions),
  ...
end type
\end{lstlisting}
\end{multicols}
Consider for instance the definition of one of the mandatory extensions in TLS 1.3, the supported version extension.
We want to model a supported version extension for the protocol ``TLS 1.2'' in LNT.
To do so, we need an extension with the supported version type, and with an extension data using the constructor ``SupportedVersions'', which takes as a parameter a supportedVersion, i.e., a list of protocol versions.
\begin{multicols}{3}
\begin{lstlisting} [language=LNT]
Extension(
  supported_version, 
  SupportedVersion ({TLS12})
)
\end{lstlisting}
\columnbreak
\begin{lstlisting} [language=LNT]
type SupportedVersions is
  list of ProtocolVersion
end type
\end{lstlisting}
\columnbreak
\begin{lstlisting} [language=LNT]
type ProtocolVersion is
  TLS12,
  ...
end type
\end{lstlisting}
\end{multicols}

\subsection{Handshake interactions}
\label{subsec:process}
Our modeling of the communication between client and server is based on the state machines (Figure \ref{fig:statemachines}) and the handshake TLS 1.3 requirements discussed in Section \ref{subsec:interactions}.

The server and the client are modeled by two processes (Appendix \ref{ap:process}) communicating by \emph{rendezvous} on gates, i.e., the communication is blocked by both sending and receiving messages: the one waiting for a rendezvous is suspended and terminates immediately after the rendezvous takes place.
Concretely, the server and the client processes correspond to their respective states machines (Figure \ref{fig:statemachines}) extended with the management of the {\tt Alert} message.
Each kind of handshake message is implemented in a process, and two additional processes by kind of handshake message sent by the client and the server.

The difficulty was the extraction of definitions from the informal definition of the TLS 1.3 handshake requirements in \cite{draft-ietf-tls-tls13-21}.
Since this informal specification is not self-contained, it refers to many documents, e.g., the alert management.

The alert management is incharge of on handling handshake errors.
If a handshake requirement is not respected, the handshake should be aborted with an alert message.
We defined the following \emph{AlertType}, an enumeration of all possible alert messages.
Each process takes as parameter an alert type, which is initialized by an ``undefined'' alert in the main process.
If a handshake requirement is not respected in a process, this one should assign the corresponding alert type to the out alert parameter, and the handshake is aborted with the corresponding alert message.
\begin{lstlisting} [language=LNT]
type AlertType is
  missing_extension,
  unexpected_message,
  unsupported_certificate,
  ...
  undefined
end type
\end{lstlisting}
During the implementation process of the handshake interactions, we took advantage of the semantics of the LNT. 
Consider for instance the following requirement: the TLS 1.3 handshake refuses renegotiation without a {\tt hello retry request} message.
The {\tt client hello} message can only arrive at the beginning of the handshake, or right after a {\tt hello retry request} message. In all other cases if a {\tt client hello} message arrives, the handshake should be aborted with an alert. 
To implement this requirement we used the LNT operator ``disrupt'', which allows at any time a possible disruption of a block by another block. 
In the following LNT code, we have for instance the possible disruption of a ``content'' behaviour by a {\tt client hello} message, followed by an alert.
\begin{lstlisting} [language=LNT]
disrupt
   ... content
by
  -- TLS 1.3 refuses renegotiation without a Hello Retry Request
  ClientHello [clientHello_c] (false, !?CH_p, HRR_P, ?alert);
  alert := unexpected_message;
  -- abort the handshake with an "unexpected_message" alert
  alert_c (alert)
end disrupt
\end{lstlisting}

\section{Validation}
\label{sec:validation}
To validate our formal model of the TLS handshake, we follow an approach based on model-based testing \cite{BroyJonsson05}.
Model-based testing enables us to corroborate a model (M) of a system under test (SUT) and an implementation of the SUT by checking the conformance between them.
Conformance testing consists of extraction of test cases (TCs) from M in order to observe whether the SUT is conform to M or not.
In this work we used TESTOR \cite{marssomateescuservwe}, a recent tool for on-the-fly conformance test case generation guided by test purposes, developed on top of the CADP toolbox \cite{garavellangmateescuserwe}.
Since there is no available implementation of the TLS 1.3 handshake as far as we know, our SUT in this validation process is an implementation of TLS 1.2 \cite{openssl}.

\subsection{Approach overview}

\begin{figure}[h!]
\centering{
\includegraphics[scale=0.27]{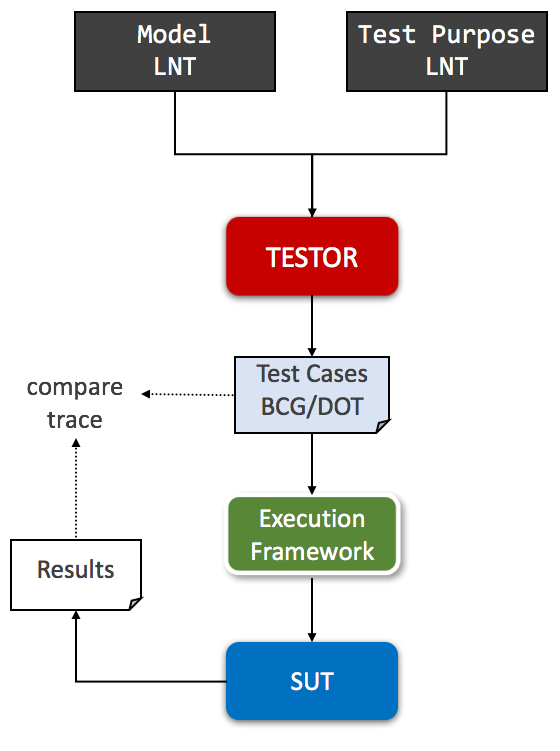}}
\caption{Validation approach of the TLS handshake}
\label{fig:approach}
\end{figure}

Our approach to validate our model is depicted in Figure~\ref{fig:approach}.
Concretely, given a formal LNT model M of the TLS handshake and a test purpose in LNT, TESTOR automatically generates a test case (TC) encoded in the BCG file format.
The generated test case has a verdict. There are three possible verdicts: \emph{fail} when the SUT is not conform to M, \emph{pass} when the test purpose is reached, or \emph{inconclusive} when there is no error but the test purpose is not reached. 
Using the CADP verification toolbox \cite{garavellangmateescuserwe}, the BCG test case is translated into a readable DOT representation, which is read by the execution framework. 
The test implementation parses the input file and categorizes the extracted data. 
In turn, this is used to make concrete Java based TLS messages for testing the SUT.

First a trace is defined according to the state transitions from the TC. Then a SUT is tested and during execution the framework creates and submits TLS messages to the server and reads the resulting output. All exchanged messages are tracked as well as their order and finally, the obtained sequence is checked for inclusion in the initial LNT model. This establishes whether the execution was conform to the initial model or not.

\subsection{Test purposes}
\label{subsec:atc}
\begin{figure}
\begin{subfigure}{0.33\textwidth}
\includegraphics[scale=0.4]{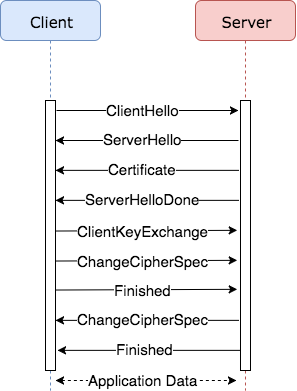} 
\caption{TC I: TLS handshake\\ with classical TLS 1.3 order}
\label{fig:test0}
\end{subfigure}
\begin{subfigure}{0.33\textwidth}
\includegraphics[scale=0.4]{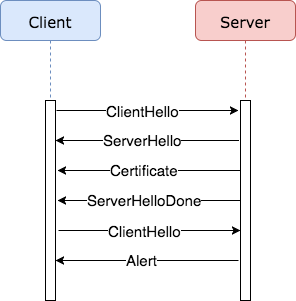} 
\caption{TC II: TLS handshake aborted\\ with an unexpected Alert}
\label{fig:test1}
\end{subfigure}
\begin{subfigure}{0.33\textwidth}
\includegraphics[scale=0.4]{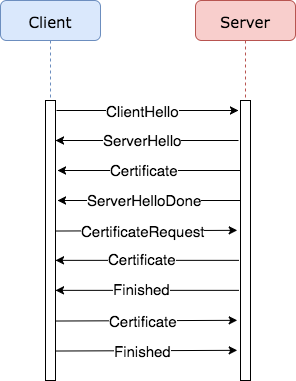}
\caption{TC III: TLS handshake\\ with renegotiation}
\label{fig:test2}
\end{subfigure}
 
\caption{Abstract test cases generated from the formal model M and the test purposes (I - III).}
\label{fig:image2}
\end{figure}

A test purpose aims to select a functionality to be tested, by guiding the selection of test cases.
We defined three test purposes corresponding to three requirements from the draft TLS 1.3 handshake \cite{draft-ietf-tls-tls13-21}:
\begin{enumerate}[I.]
  \item The protocol messages must be sent in the right order, using classical TLS 1.3 order, (without {\tt hello retry request} message).
  \item The handshake must be aborted with an ``unexpected_message'' alert, if there is a client renegotiation.
  \item The protocol messages are sent in the right order with an incorrect key shared (with {\tt hello retry request} message).
\end{enumerate}
The LNT models of these test purposes are given in Appendix \ref{ap:tp1}, \ref{ap:tp2}, \ref{ap:tp3}, respectively.
Given our formal LNT model of the TLS handshake and our test purposes, we automatically generated with TESTOR the abstract test cases, represented in a simplified form in Figure \ref{fig:test0}, Figure \ref{fig:test1}, and Figure \ref{fig:test2}.

During this validation process we found an error in our model, with test purpose III.
We couldn't reach this accepting state of purpose, because of our intial  implementation  of this requirement:
``The {\tt server hello} message should have the same cryptographic information as the {\tt hello retry request} message.''
We change our model to correct this issue.
In fact, we were assigning to the {\tt hello retry request} encryption data the value of the {\tt server hello} encryption data, whereas the {\tt hello retry request} message arrives before the {\tt server hello} message.

The generated abstract test cases are then refined for becoming suitable inputs of the execution framework on the SUT. This described in the next section.

\subsection{Test execution}
\label{sec:texec}

If a client wants to establish a communication channel over HTTPS with the server, then the TLS handshake procedure is done automatically. However, for testing purposes it is necessary to emulate this interaction in a controlled way. In fact, we want to establish a connection to a TLS implementation with our program and automatically test the SUT by following a formal specification. For this sake, we use TLS-Attacker \cite{tlsattacker}, which already encompasses a basic functionality in order to communicate with a SUT. We extend the tool with additional functionality and establish a connection with the formal model. 

First, the representation of the test case encoded in the DOT format is parsed and categorized by the tool. Table~\ref{table:dot} depicts the obtained results for the TC III (Figure \ref{fig:test2}) and instructs the execution. Then, the framework creates TLS messages on the fly according to the table, submits them against a SUT, and reads its responses. Since no concrete values for the parameters of the messages are assigned, some default values from the tool are used for testing purposes. Finally, the execution terminates, either after the sequence from the table has been executed, or if interruptions have occurred. In any case, the resulting trace will give evidence about the execution. 

{\small
\begin{table}
\caption{Extracted transitions from the DOT representation of TC III} 
\centering 
\begin{tabular}{|c |c |c |} 
\hline
Pre & Action & Post  \\ [0.5ex] 
\hline\hline
0 & CLIENTHELLO & 1  \\
1 & SERVERHELLO & 2 \\
2 & CERTIFICATE_S & 3 \\
3 & SERVERHELLODONE & 4 \\
4 & CERTIFICATEREQUEST & 5 \\ 
5 & CERTIFICATE\_S & 6 \\ 
6 & FINISHED\_S & 7\\ 
7 & CERTIFICATE\_C & 8 \\ 
8 & FINISHED\_C & 9 \\ 
9 & exit & 10 \\ 
\hline\hline
\end{tabular} 
\label{table:dot}
\end{table}
} 
The test oracle in this case represents the initial list of expected states derived from the test case. After every executed action the table gives an expected output in that state. Any discrepancy means that the test did not proceed as expected. 

\subsection{Analysis of test execution}
\label{sec:eval}

For the evaluation two different test cases have been generated. The framework tests OpenSSL version~1.0.1e \cite{openssl} and tracks the interaction. The generated abstract test cases from the TLS 1.3 handshake model (Figure~\ref{fig:test1} and Figure~\ref{fig:test2}) are converted into abstract test cases for the execution framework. As already mentioned, the generated test cases from LNT are parsed by the implementation. The tool handles the concretization automatically by itself. 

One of the ideas behind our approach is applicability, i.e, it should be possible to test a wide range of TLS implementations by only slightly manipulating the overall system. For example, only the port has to be changed manually in order to test another application. The test results give information about a verdict. The test case can be either complete, in case that all paths inside the model have been traversed, or incomplete. The second scenario can occur either due to unexpected behavior because of concrete values inside the TLS' parameters.

In general, since TLS is very complex and manipulates a large number of concrete values, the individual parameter assignments do have a big impact on the execution. However, in this work we will omit the test case concretization possibilities.

When testing in accordance to the TLS 1.3 handshake LNT formal model M, we achieve a complete test run. First, the initial {\tt clientHello} is sent, after which the standard three server-side responses occur. Then we submit an unexpected {\tt clientHello} again and try to re-negotiate the handshake procedure. As expected, the server sends an Alert message with the following description in TLS-Attacker: 

{\small\begin{verbatim}
ALERT message:
  Level: FATAL
  Description: UNEXPECTED_MESSAGE
\end{verbatim}}

The trace obtained from the execution framework confirms that all TLS messages have been sent or received. The system responded as expected when being confronted with unexpected input. Thus, the behavior of the SUT is in conformance to the given TLS 1.3 handshake LNT formal model. We conclude that the test case has been successful.

{\small\begin{verbatim}
Action #1: CLIENT_HELLO
Action #2: SERVER_HELLO
Action #3: CERTIFICATE
Action #4: SERVER_HELLO_DONE
Action #5: CLIENT_HELLO
Action #6: ALERT
\end{verbatim}}

For the TC III (Figure \ref{fig:test2}), which corresponds to Table~\ref{table:dot}, an additional client-side {\tt CertificateRequest} was specified. Since sending this TLS event is not mandatory, we want to check whether the execution might proceed further or ignore the message. However, the obtained trace indicates that, at least the current version of OpenSSL, does not reply to the request with the expected certificate. 

{\small\begin{verbatim}
Action #1: CLIENT_HELLO
Action #2: SERVER_HELLO
Action #3: CERTIFICATE
Action #4: SERVER_HELLO_DONE
Action #5: CERTIFICATE_REQUEST
Action #6: ALERT
\end{verbatim}}

Just the opposite is the case wherethe server replies with an error and closes the connection:

{\small\begin{verbatim}
routines:ACCEPT_SR_KEY_EXCH:unexpected message
\end{verbatim}}

The verdict here is an incomplete test case. This is likely to be caused by the concrete implementation of OpenSSL. Either a {\tt CertificateRequest} is not tolerated during this point of the handshake, or a preceding concrete value causes the issue at this point. According to given circumstances, we conclude that the SUT does not behave in conformance to the model.  

\section{Related Work}
\label{sec:rel}

\paragraph{Formal specification of TLS:}
In \cite{poll15} the authors propose a technique for derivation of models from the specification of TLS implementations. They focus on session languages for sequences of actions from a protocol and automatically infer formal specifications of these languages. Their approach is a black-box testing-based approach where statecharts are extracted from implementations. In this way security leaks might be detected in the host application. As opposed, in our approach specifications are not extracted but already given.
\cite{diaz04} discusses formal methods for the TLS handshake protocol for e-commerce properties. They rely on the tool UPPAAL for verifying the TLS functionalities. They describe the handshake and the behavior of TLS messages in forms of timed automata. Then, they validate the protocol by checking the correct message flow between client and server. 

\paragraph{Conformance testing of TLS:}
A set of tools for TLS 1.3, which includes a conformance testing technique, is provided \cite{km16}. The TLS conformance checker communicates with nqsb-TLS, a TLS implementation, whereas a test oracle tracks a session from the interaction and checks whether it is conform with the protocol specification. Then, these sessions can be replicated and run against another TLS under test. A different behavior of the SUT in comparison with nqsb-TLS indicates a discrepancy. The main difference to our approach is that the authors rely on a reference implementation of conformance checks. On the contrary, our approach compares the outputs of a SUT with test cases obtained from a formal LNT model. 

\paragraph{White box testing:}
The authors of \cite{affeldt13} present an approach with the goal to improve TLS implementations. For this sake, a framework for verification of C implementations for TLS functionality is provided. Formal definitions are provided for TLS in order to ease the verification of implementations. Here, a specification of TLS event packets is provided with corresponding extensions in the clientHello message. The authors use their framework for verification of functions from PolarSSL, a TLS implementation. In contrast to their work, our approach relies on LNT for formalization of TLS. Also, the testing process follows a black-box conformance testing approach instead of individual C function verification.

\section{Conclusion and Future Work}
\label{sec:conc}

In this paper we have presented a formal LNT model of the draft Transport Layer Security TLS handshake protocol version 1.3, the first formal model as far as we know. 
We first gave an overview of this widely-used security protocol, then we discussed  the choices and the challenges of this implementation in LNT. 
Finally, we validated our model using conformance testing techniques, and discussed the results.
As outlined in other security testing related works (e.g., \cite{ictss17}), multiple TLS implementations behave differently when being confronted with the same inputs. Whereas some applications indicate a high degree of resistance against any unexpected messages in terms of abstract or concrete values, others are more tolerant. This leads to the conclusion that TLS implementations do not always follow the strict specification of the protocol. In this case, conformance testing can help in order to detect the discrepancies.

In the future, we plan to enhance our TLS.1.3 handshake LNT model, to handle more extensions, and to implement the {\tt end of early data}, the {\tt new session ticket} and {\tt key update} messages.
Our model is easily extendable, thanks to the ``union like'' types.  
We also plan to improve our validation process by using as a system under test the future implementations of the TLS 1.3 handshake and by specifying TLS attacks as test purposes.

\section*{Acknowledgment}
We are grateful to Hubert Garavel and Wendelin Serwe for helpful remarks about the paper and the LNT model.
The research presented in the paper has been funded in part by the Austrian Research Promotion Agency (FFG) under grant 851205 (Security ProtocoL Interaction Testing in Practice - SPLIT), RIDINGS (RIgorous DesIgN of GALS Systems) project of the PHC Amadeus program, and the R\'{e}gion Auvergne-Rh\^{o}ne-Alpes within the program ARC 6.

\bibliographystyle{eptcs}
\bibliography{related}

\clearpage
\appendix
\section{Formal Model of the TLS 1.3 Handshake in LNT}
\label{ap:model}
Our model is decomposed in three modules, a first one with the type definitions, 
a second one with process definitions common to specification and the test purposes, and a last one with the process definitions used only in the specification.
The decomposition avoids the duplication of code during the validation process. 
\subsection{Handshake types}
\begin{lstlisting}[language=LNT, firstline=1, lastline=882]
module HANDSHAKETYPE
with "==", "!=", "get", "set", "head", "tail"
is

-------------------------------------------------------------------------------

type Random32 is
  28byteRand,
  undefined
end type

-------------------------------------------------------------------------------

type ProtocolVersion is
  TLS10,
  TLS11,
  TLS12,
  TLS13,
  DTLS10,
  DTLS12,
  T_NULL
end type

-------------------------------------------------------------------------------

type AlertType is
  -- cf. Section 6 of draft-ietf-tls-tls13-24
  bad_certificate,
  certificate_required,
  decode_error,
  decrypt_error,
  illegal_parameter,
  insufficient_security,
  missing_extension,
  protocol_version,
  unexpected_message,
  unsupported_certificate,
  undefined
end type

-------------------------------------------------------------------------------

type CipherType is
  -- cf. Appendix B.4 of draft-ietf-tls-tls13-24
  TLS_AES_128_GCM_SHA256,
  TLS_AES_256_GCM_SHA384,
  TLS_CHACHA20_POLY1305_SHA256,
  TLS_AES_128_CCM_SHA256,
  TLS_AES_128_CCM_8_SHA256
end type

-------------------------------------------------------------------------------

type SignatureScheme is
  -- cf. Section 4.3 of draft-ietf-tls-tls13-24
  -- RSASSA-PKCS1-v1_5 algorithms
  rsa_pkcs1_sha256,
  rsa_pkcs1_sha384,
  rsa_pkcs1_sha512,
  -- ECDSA algorithms
  ecdsa_secp256r1_sha256,
  ecdsa_secp384r1_sha384,
  ecdsa_secp521r1_sha512,
  -- RSASSA-PSS algorithms
  rsa_pss_sha256,
  rsa_pss_sha384,
  rsa_pss_sha512,
  -- EdDSA algorithms
  ed25519,
  ed448,
  -- Legacy algorithms
  rsa_pkcs1_sha1,
  ecdsa_sha1,
  -- Reserved Code Points
  private_use,
  unknown
end type

-------------------------------------------------------------------------------

type Compression_methods is
  T_NULL,
  DEFLATE,
  LZS
end type

-------------------------------------------------------------------------------

type Session_id is
  T_NULL,
  32byteID
end type

-------------------------------------------------------------------------------

type Ciphers is
  list of ciphertype
end type

-------------------------------------------------------------------------------

type ExtensionType is
  -- cf. Section 4.2 of draft-ietf-tls-tls13-24
  server_name,
  max_fragment_length,
  status_request,
  supported_groups,
  signature_algorithms,    -- signatureScheme
  use_srtp,
  heartbeat,
  application_layer_protocol_negotiation,
  signed_certificate_timestamp,
  client_certificate_type,
  server_certificate_type,
  padding,
  key_share,
  pre_shared_key,
  early_data,
  supported_versions,
  cookie,
  psk_key_exchange_modes,
  certificate_authorities,
  void_filters,
  post_handshake_auth
end type

-------------------------------------------------------------------------------

type BYTE is
  x00, x01, x02, x03, x04, x05, x06, x07, x08, x09, x0F, xFF
end type

-------------------------------------------------------------------------------

type Cookie is
  list of BYTE
end type
-------------------------------------------------------------------------------

type CertificateAuthoritiesExtension is
  list of BYTE
end type

-------------------------------------------------------------------------------

type CertificateExtensionOid is
  list of BYTE
end type

-------------------------------------------------------------------------------

type CertificateExtensionValues is
  list of BYTE
end type

-------------------------------------------------------------------------------

type CertificateType is
  RawPublicKey,
  X509
end type

-------------------------------------------------------------------------------

type NamedGroup is
  -- cf. Appendix B.3.1.4 of draft-ietf-tls-tls13-24
  -- Elliptic Curve Groups (ECDHE)
  ecp256r1,          -- (0x0017),
  secp384r1,         -- (0x0018),
  secp521r1,         -- (0x0019),
  x25519,            -- (0x001D),
  x448,              -- (0x001E),
  -- Finite Field Groups (DHE)
  ffdhe2048,         -- (0x0100),
  ffdhe3072,         -- (0x0101),
  ffdhe4096,         -- (0x0102),
  ffdhe6144,         -- (0x0103),
  ffdhe8192,         -- (0x0104),
  -- Reserved Code Points
  ffdhe_private_use, -- (0x01FC..0x01FF),
  ecdhe_private_use, -- (0xFE00..0xFEFF),
  unknown            -- (0xFFFF)
end type

-------------------------------------------------------------------------------

type NamedGroupList is
  list of NamedGroup
end type

-------------------------------------------------------------------------------

type KeyShareEntry is
  list of NamedGroup
end type

-------------------------------------------------------------------------------

type OIDFilter is
  peer (oid: CertificateExtensionOid, value: CertificateExtensionValues)
end type

-------------------------------------------------------------------------------

type OIDFilterExtension is
  list of OIDFilter
end type

-------------------------------------------------------------------------------
type SignatureSchemeList is
  -- supported_signature_algorithms
  list of SignatureScheme
end type
-------------------------------------------------------------------------------

type SupportedVersions is
  list of ProtocolVersion
end type

-------------------------------------------------------------------------------
type ExtensionData is
  Cookie (c: Cookie),
  CertificateAuthoritiesExtension (cae: CertificateAuthoritiesExtension),
  CertificateType (ct: CertificateType),
  CertificateStatusRequest (csr: CertificateStatusRequest),
  KeyShareEntry (kse: KeyShareEntry),
  NamedGroupList (ng: NamedGroupList),
  OIDFilterExtension (oidfe: OIDFilterExtension),
  SignatureSchemeList (ss: SignatureSchemeList),
  SupportedVersions (sv: SupportedVersions)
end type

-------------------------------------------------------------------------------

type Extension is
  Extension (extension_type: ExtensionType, extension_data: ExtensionData)
end type

-------------------------------------------------------------------------------

type Extensions is
  list of Extension
with "reverse", "member", "remove"
end type

-------------------------------------------------------------------------------

type ResponderIdList is
  list of BYTE
end type

-------------------------------------------------------------------------------

type OCSPStatusRequest is
  OCSPStatusRequest (responderId: ResponderIdList, extensions: Extensions)
end type

-------------------------------------------------------------------------------

type CertificateStatusRequest is
  255,
  OCSP (ocsp: OCSPStatusRequest)
end type

-------------------------------------------------------------------------------

type Entry is
  Entry (entry_type: certificateType, extensions: Extensions)
end type

-------------------------------------------------------------------------------

type Entries is
 list of Entry
with "reverse", "member", "remove"
end type

-------------------------------------------------------------------------------

type CertificateRequestContext is
  list of BYTE
end type

-------------------------------------------------------------------------------

type Signature is
  list of BYTE
end type

-------------------------------------------------------------------------------

type MAC is
  server_handshake_traffic_secret,
  client_handshake_traffic_secret,
  client_application_traffic_secret
end type

-------------------------------------------------------------------------------

type ClientHello is
  ClientHello (legacy_version: ProtocolVersion,
               random: Random32,
               legacy_session_id: Session_id,
               cipher_suite: ciphers,
               legacy_compression_methods: Compression_methods,
               extensions: Extensions)
end type

-------------------------------------------------------------------------------

type ServerHello is
  ServerHello (protocol_version: ProtocolVersion,
               random: Random32,
               cipher_suite: ciphers,
               extensions: Extensions)
end type

-------------------------------------------------------------------------------

type HelloRetryRequest is
  HelloRetryRequest (protocol_version: ProtocolVersion,
                     cipher_suite: ciphers,
                     extensions: Extensions)
end type

-------------------------------------------------------------------------------

type EncryptedExtensions is
  EncryptedExtensions (extensions: Extensions)
end type

-------------------------------------------------------------------------------

type CertificateRequest is
  CertificateRequest (certificate_context: CertificateRequestContext,
                      extensions: Extensions)
end type

-------------------------------------------------------------------------------

type Certificate is
  Certificate (crc: certificateRequestContext, certificate_list: Entries)
end type

-------------------------------------------------------------------------------

type CertificateVerify is
  CertificateVerify (algorithm: SignatureSchemeList, ss: Signature)
end type

-------------------------------------------------------------------------------

type Finished is
  Finished (hmac: MAC)
end type

-------------------------------------------------------------------------------

type EndOfEarlyData is
  EndOfEarlyData (b: BOOL)
end type

------------------------------------------------------------------------------
-- Constants
------------------------------------------------------------------------------

function random_signature_algorithms: SignatureSchemeList is
  return {rsa_pkcs1_sha256, rsa_pkcs1_sha384,
            rsa_pkcs1_sha512, ecdsa_secp256r1_sha256}
end function

------------------------------------------------------------------------------

function random_select_ciphers : ciphers is
  return {TLS_AES_128_GCM_SHA256, TLS_AES_256_GCM_SHA384,
            TLS_CHACHA20_POLY1305_SHA256, TLS_AES_128_CCM_SHA256}
end function

------------------------------------------------------------------------------

function random_certificate_authorites : CertificateAuthoritiesExtension is
  return {xFF}
end function

------------------------------------------------------------------------------
-- Functions
------------------------------------------------------------------------------

function is_extension (in ex: ExtensionType, in var exts: Extensions): BOOL is
  loop
    if exts == {} then
      return false
    elsif head (exts).extension_type == ex then
      return true
    else
      exts := tail (exts)
    end if
  end loop
end function

-------------------------------------------------------------------------------

function cp_extension (in ex: ExtensionType, in var exts: Extensions): Extension is
  var list_ext: Extensions, e: Extension in
    loop
      assert exts != {};
      e := head (exts);
      if e.extension_type == ex then
        return e
      else
        exts := tail (exts)
      end if
    end loop
  end var
end function

-------------------------------------------------------------------------------

function add_extension (ex: Extension, exts: Extensions): Extensions is
  if is_extension (ex.extension_type, exts) == false then
    return cons (ex, exts)
  else
    return exts
  end if
end function

-------------------------------------------------------------------------------

function remove_type_extension (in ex: ExtensionType,
                                in var exts: Extensions): Extensions is
  var t_list: Extensions, e: Extension in
    t_list := {};
    loop
      if exts == {} then
        return t_list
      end if;
      e := head (exts);
      if e.extension_type != ex then
        t_list := cons (e, t_list)
      end if;
      exts := tail (exts)
    end loop
  end var
end function

-------------------------------------------------------------------------------

function add_Entry (certificate_type: CertificateType,
                    CH_p: ClientHello) : Entry is
  var exts: Extensions in
    if is_extension (status_request, CH_p.extensions) then
      exts := {cp_extension (status_request, CH_p.extensions)}
    elsif is_extension (signed_certificate_timestamp, CH_p.extensions) then
      exts := {cp_extension (signed_certificate_timestamp, CH_p.extensions)}
    else
      exts := {}
    end if;
    return Entry (certificate_type, exts)
  end var
end function

end module
\end{lstlisting}
\label{ap:types}
\clearpage
\subsection{Common handshake interactions}
\begin{lstlisting}[language=LNT]
module handshake_interactions (HANDSHAKETYPE) is

-------------------------------------------------------------------------------

channel CH is (ClientHello) end channel

channel SH is (ServerHello) end channel

channel HRR is (HelloRetryRequest) end channel

channel EE is (EncryptedExtensions) end channel

channel CR is (CertificateRequest) end channel

channel C is (Certificate) end channel

channel CV is (CertificateVerify) end channel

channel F is (Finished) end channel

channel EOED is (EndOfEarlyData) end channel

channel A is (AlertType) end channel

-------------------------------------------------------------------------------

process ServerHello [serverh: SH] is
  serverh (ServerHello(TLS12, 28byteRand, random_select_ciphers, {}))
end process

-------------------------------------------------------------------------------

process HelloRetryRequest [hellorr: HRR] (in out HRR_p: HelloRetryRequest) is
  HRR_p := HRR_p.{protocol_version => TLS12,
                  cipher_suite => random_select_ciphers,
                  extensions => {}};
  hellorr (HelloRetryRequest (HRR_p.protocol_version,
                              HRR_p.cipher_suite, HRR_p.extensions))
end process

-------------------------------------------------------------------------------

process CertificateVerify_S [certificatev: CV] is
  certificatev(CertificateVerify ({rsa_pkcs1_sha256}, {}))
end process

-------------------------------------------------------------------------------

process Finished_S [finished: F] is
  finished (Finished (server_handshake_traffic_secret))
end process

-------------------------------------------------------------------------------

process Finished_C [finished: F] is
  finished (Finished (client_handshake_traffic_secret))
end process

end module
\end{lstlisting}

\clearpage
\subsection{Handshake}
\begin{lstlisting}[language=LNT]
module handshake (handshake_interactions) is

-------------------------------------------------------------------------------

process ClientHello [clienth: CH] (is_hello_retry_request: bool,
                                   in out CH_p: ClientHello,
                                   HRR_p: HelloRetryRequest,
                                   out alert: AlertType) is
  if is_hello_retry_request then
    -- Hello retry request sent by the server
    if (CH_p.legacy_version == HRR_p.protocol_version) and
       (CH_p.cipher_suite == HRR_p.cipher_suite) and
       (CH_p.extensions == HRR_p.extensions) then
      -- The request would not result in any change in the ClientHello
      alert := illegal_parameter
    else
      if is_extension (early_data, CH_p.extensions) then
        CH_p := CH_p.{extensions =>
                        remove_type_extension (early_data, CH_p.extensions)}
      end if;
      alert := undefined
    end if
  else
    -- First ClientHello message
    -- initialisation with standard arguments
    CH_p := CH_p.{legacy_version => TLS12,
                  random => 28byteRand,
                  legacy_session_id => T_NULL,
                  -- AED algo/HKDF hash pairs supported by the Client
                  cipher_suite => {},
                  legacy_compression_methods => T_NULL,
                  extensions => {Extension (supported_versions,
                                            SupportedVersions ({TLS13}))}};
    select
      -- desire the server to authenticate via a certificate
      var sas: SignatureSchemeList in
        sas := random_signature_algorithms;
        CH_p := CH_p.{extensions =>
                        add_extension (Extension (signature_algorithms,
                                                  SignatureSchemeList(sas)),
                                       CH_p.extensions)}
      end var
    []
      -- client MAY send the "certificate_authorities"
      var cr: CertificateAuthoritiesExtension in
        cr := random_certificate_authorites;
        CH_p := CH_p.{extensions =>
                        add_extension
                          (Extension (certificate_authorities,
                                      CertificateAuthoritiesExtension(cr)),
                           CH_p.extensions)}
      end var
    []
      CH_p := CH_p.{cipher_suite => random_select_ciphers}
    end select;
    alert := undefined
  end if;
  clienth (ClientHello (CH_p.legacy_version, CH_p.random,
                        CH_p.legacy_session_id, CH_p.cipher_suite,
                        CH_p.legacy_compression_methods, CH_p.extensions))
end process

-------------------------------------------------------------------------------

process EncryptedExtensions [encryptede: EE] is
  var v_extensions: Extensions in
    select
      v_extensions := {}
    []
      -- forbidden extensions
      v_extensions := {Extension (certificate_authorities,
                                  CertificateAuthoritiesExtension
                                    ({x00, xFF}))}
    end select;
    encryptede (EncryptedExtensions (v_extensions))
  end var
end process

-------------------------------------------------------------------------------

process CertificateRequest [certificater: CR] is
  var v_certificate_context: CertificateRequestContext,
      v_extensions: Extensions in
    v_certificate_context := {};
    v_extensions := {Extension (void_filters,
                                OIDFilterExtension ({peer ({x00}, {x00})}))};
    select
      null
    []
      -- Server MAY send the "certificate_authorities"
      var cr: CertificateAuthoritiesExtension in
        cr := random_certificate_authorites;
        v_extensions :=
          add_extension (Extension (certificate_authorities,
                                    CertificateAuthoritiesExtension (cr)),
                         v_extensions)
      end var
    end select;
    v_extensions :=
      add_extension (Extension (signature_algorithms,
                                SignatureSchemeList ({rsa_pkcs1_sha256})),
                     v_extensions);
    certificater (CertificateRequest (v_certificate_context, v_extensions))
  end var
end process

-------------------------------------------------------------------------------

process Certificate_S [certificate: C]
                      (in out C_p: Certificate,
                       CH_p: ClientHello, CR_p: CertificateRequest) is
  var ed: ExtensionData in
    C_p := C_p.{crc => CR_p.certificate_context};
    if is_extension (server_certificate_type, CH_p.extensions) then
      var ex: Extension in
        ex := cp_extension (server_certificate_type, CH_p.extensions);
        ed := ex.extension_data;
        if ed.ct != RawPublicKey then
          -- X509
          ed := CertificateType (x509)
        end if
      end var
    else
      -- X509
      ed := CertificateType (x509)
    end if;
    C_p := C_p.{certificate_list =>
                  cons (add_Entry (ed.ct, CH_p), C_p.certificate_list)};
    certificate (Certificate(C_p.crc, C_p.certificate_list))
  end var
end process

-------------------------------------------------------------------------------

process Certificate_C [certificate: C, certificatev: CV]
                      (in out C_p: Certificate, CH_p: ClientHello,
                      CR_p: CertificateRequest) is
  -- Certificate available
  if is_extension (client_certificate_type, CH_p.extensions) then
    -- Response to a CertificateRequest
    C_p := C_p.{crc => CR_p.certificate_context};
    var ex: Extension, ed: ExtensionData in
      ex := cp_extension (client_certificate_type, CH_p.extensions);
      ed := ex.extension_data;
      if ed.ct != RawPublicKey then
        -- X509
        ed := CertificateType (x509)
      end if;
      C_p := C_p.{certificate_list => 
                    cons (add_Entry(ed.ct, CH_p), C_p.certificate_list)}
    end var;
    CertificateVerify_C [certificatev]
  else
    -- No suitable certificate is available
    -- Client MUST send a Certificate message containing no certificates
    C_p := C_p.{certificate_list => {}, crc => {}}
  end if;
  certificate (Certificate(C_p.crc, C_p.certificate_list))
end process

-------------------------------------------------------------------------------

process CertificateVerify_C [certificatev: CV] is
  certificatev(CertificateVerify ({rsa_pkcs1_sha256}, {}))
end process

-------------------------------------------------------------------------------

process EndOfEarlyData [endOEdata: EOED] is
  endOEdata (EndOfEarlyData (true))
end process

-------------------------------------------------------------------------------

process Client [clientHello_c: CH, serverHello_c: SH,
                helloRetryRequest_c: HRR, encryptedExtensions_c: EE,
                certificateRequest_c: CR, certificate_c_c, certificate_s_c: C,
                certificateVerify_c_c, certificateVerify_s_c: CV,
                finished_c_c, finished_s_c: F, endOfEarlyData_c: EOED,
                alert_c: A] is
  var alert: AlertType, CH_p: CLientHello, C_C_p: Certificate,
      CR_p: CertificateRequest, HRR_P: HelloRetryRequest,
      is_earlyData, is_helloRequest, is_PSK: BOOL in
    -- Initialisation of variables
    CH_p := ClientHello (T_NULL, undefined, T_NULL, {}, T_NULL, {});
    HRR_P := HelloRetryRequest (T_NULL, {}, {});
    CR_p := CertificateRequest ({},{});
    C_C_p := Certificate ({},{});
    is_earlyData := false;
    is_helloRequest := false;
    is_PSK := false;
    -- A. Start
    loop L in
      -- client key exchange [K_send = early data]
      ClientHello [clientHello_c] (is_helloRequest, !?CH_p, HRR_P, ?alert);
      if alert != undefined then
        -- abort the handshake with an alert
        alert_c (alert)
      else
        -- B. WAIT_ServerHello
        select
          helloRetryRequest_c (?HRR_P);
          is_helloRequest := true
        []
          serverHello_c (?any ServerHello);
          break L
        []
          -- protocol messages sent in the wrong order
          select
            encryptedExtensions_c (?any EncryptedExtensions)
          []
            certificateRequest_c (?any CertificateRequest)
          []
            certificate_s_c (?any Certificate)
          []
            certificateVerify_s_c (?any CertificateVerify)
          []
            finished_s_c (?any Finished)
          end select;
          alert := unexpected_message;
          -- abort the handshake with an "unexpected_message" alert
          alert_c (alert)
        end select
      end if
    end loop;
    if alert == undefined then
      -- K_recv = handshake
      -- C. WAIT_EncryptedExtention
      -- Recv EncryptedExtensions
      encryptedExtensions_c (?any EncryptedExtensions);
      select
        -- Using certificate
        -- D. WAIT_CERT_CR
        select
          -- Recv CertificateRequest
          certificateRequest_c (?CR_p);
          -- E. WAIT_Certificate
          certificate_s_c (?any Certificate)
        []
          certificate_s_c (?any Certificate)
        end select;
        -- F. Wait CertificateVerify
        certificateVerify_s_c (?any CertificateVerify)
      []
        -- using PSK
        is_PSK := true
      end select;
      -- G. Wait_Finished
      finished_s_c (?any Finished);
      if is_earlyData then
        -- [Send EndOfEarlyData]
        EndOfEarlyData [endOfEarlyData_c]
      end if;
      -- K_send = handshake
      -- [Send Certificate [+ CertificateVerify]]
      if is_PSK != true then
        -- client authentification
        Certificate_C [certificate_c_c, certificateVerify_c_c]
                      (!?C_C_p, CH_p, CR_p);
        use C_C_p
      end if;
      -- Send finished
      Finished_C [finished_c_c]
      -- K_send = K_recv = application
      -- E. Connected
    end if
  end var
end process

-------------------------------------------------------------------------------

process Server [clientHello_c: CH, serverHello_c: SH,
                helloRetryRequest_c: HRR, encryptedExtensions_c: EE,
                certificateRequest_c: CR, certificate_c_c, certificate_s_c: C,
                certificateVerify_c_c, certificateVerify_s_c: CV,
                finished_c_c, finished_s_c: F, endOfEarlyData_c: EOED,
                alert_c: A] is
  var alert: AlertType, CH_p: CLientHello, C_S_p, C_C_p: Certificate,
      CR_p: CertificateRequest, HRR_P: HelloRetryRequest, is_earlyData,
      is_PSK: BOOL in
    -- Initialisation of variables
    alert := undefined;
    use alert;
    HRR_P := HelloRetryRequest (T_NULL, {}, {});
    CR_p := CertificateRequest ({},{});
    C_S_p := Certificate ({},{});
    is_earlyData := false;
    is_PSK := false;
    -- A. START
    loop L in
      -- Rec ClientHello
      -- B. RECVD_CH
      clientHello_c (?CH_p);
      select
        -- right parameters received
        break L
      []
        -- protocol messages sent in the right order with incorrect key shared
        HelloRetryRequest [helloRetryRequest_c] (!?HRR_P)
      end select
    end loop;
    disrupt
      -- Select parameters
      -- C. Negotiated
      -- Send ServerHello
      ServerHello [serverHello_c];
      -- K_send = handshake
      -- Send EncryptedExtensions
      EncryptedExtensions [encryptedExtensions_c];
      if is_PSK != true then
        -- [Send CertificateRequest]
        CertificateRequest [certificateRequest_c];
        -- [Send Certificate + CertificateVerify]
        Certificate_S [certificate_s_c](!?C_S_p, CH_p, CR_p);
        use C_S_p;
        CertificateVerify_S [certificateVerify_s_c]
      end if;
      -- Send finished
      Finished_S [finished_s_c];
      -- K_send = application
      if is_earlyData then
        -- 0 - RTT
        -- K_recv = early data
        loop L3 in
          -- D. WAIT_EOED
          select
            -- Recv early data
            is_earlyData := true;
            use is_earlyData
          []
            -- Recv EndOfEarlyData
            endOfEarlyData_c (?any EndOfEarlyData);
            -- K_recv = handshake
            break L3
          end select
        end loop
      end if;
      -- E. WAIT_FLIGHT2
      if is_PSK == false then
        -- Client auth
        -- F. WAIT_CERT + G. WAIT_CV
        select
          -- G. WAIT_CV
          -- Recv CertificateRecv + CertificateVerify
          -- client authentification
          certificate_c_c (?any Certificate);
          certificateVerify_c_c (?any CertificateVerify)
        []
          -- Recv empty certificate
          certificate_c_c (?C_C_p)  where (C_C_p == Certificate ({},{}))
        end select
      end if;
      -- H. WAIT_FINISHED
      -- Recv finished
      finished_c_c (?any Finished)
      -- K_rev = application
      -- I. CONNECTED
    by
      -- TLS 1.3 refuses renegociation without a Hello Retry Request
      clientHello_c (?any CLientHello);
      alert := unexpected_message;
      -- abort the handshake with an "unexpected_message" alert
      alert_c (alert)
    end disrupt
  end var
end process

-------------------------------------------------------------------------------

process MAIN [clientHello_c: CH, serverHello_c: SH, helloRetryRequest_c: HRR,
              encryptedExtensions_c: EE, certificateRequest_c: CR,
              certificate_c_c, certificate_s_c: C, certificateVerify_c_c,
              certificateVerify_s_c: CV, finished_c_c, finished_s_c: F,
              endOfEarlyData_c: EOED, alert_c: A] is
  par
    clientHello_c, serverHello_c, helloRetryRequest_c,
    encryptedExtensions_c, certificateRequest_c, certificate_c_c,
    certificate_s_c, certificateVerify_c_c, certificateVerify_s_c,
    finished_c_c, finished_s_c, endOfEarlyData_c, alert_c ->
      Client [clientHello_c, serverHello_c, helloRetryRequest_c,
              encryptedExtensions_c, certificateRequest_c, certificate_c_c,
              certificate_s_c, certificateVerify_c_c, certificateVerify_s_c,
              finished_c_c, finished_s_c, endOfEarlyData_c, alert_c]
  ||
    clientHello_c, serverHello_c, helloRetryRequest_c,
    encryptedExtensions_c, certificateRequest_c, certificate_c_c,
    certificate_s_c, certificateVerify_c_c, certificateVerify_s_c,
    finished_c_c, finished_s_c, endOfEarlyData_c, alert_c ->
      Server [clientHello_c, serverHello_c, helloRetryRequest_c,
              encryptedExtensions_c, certificateRequest_c, certificate_c_c,
              certificate_s_c, certificateVerify_c_c, certificateVerify_s_c,
              finished_c_c, finished_s_c, endOfEarlyData_c, alert_c]
  end par

end process

end module
\end{lstlisting}
\label{ap:process}

\clearpage
\section{Test Purposes in LNT}
Our three test purposes require four additional modules: a first one with the process definitions common to all three test purposes, plus, for each of the three test purpose, one module with specific definitions.
\begin{lstlisting}[language=LNT]
module common_TP_interactions (handshake_interactions) is

------------------------------------------------------------------------------

process ClientHello_TP [clienth: CH] (is_hello_retry_request: bool,
                                      in out CH_p: ClientHello,
                                      HRR_p: HelloRetryRequest,
                                      out alert: AlertType) is
  -- Hello retry request sent by the server
  if is_hello_retry_request then
    if (CH_p.legacy_version == HRR_p.protocol_version) and
       (CH_p.cipher_suite == HRR_p.cipher_suite) and
       (CH_p.extensions == HRR_p.extensions) then
      -- The request would not result in any change in the ClientHello
      alert := illegal_parameter
    else
      alert := undefined
    end if
  else
    use is_hello_retry_request, HRR_p;
    -- initialisation with standard arguments
    CH_p := CH_p.{legacy_version => TLS12,
                  random => 28byteRand,
                  legacy_session_id => T_NULL,
                  -- AED algo/HKDF hash pairs supported by the Client
                  cipher_suite => {},
                  legacy_compression_methods => T_NULL,
                  extensions => {Extension (supported_versions,
                                            SupportedVersions ({TLS13}))}};
    var sas: SignatureSchemeList in
      sas := random_signature_algorithms;
      CH_p := CH_p.{extensions =>
                      add_extension (Extension (signature_algorithms,
                                                SignatureSchemeList(sas)),
                                     CH_p.extensions)}
    end var;
    alert := undefined
  end if;
  clienth (ClientHello (CH_p.legacy_version, CH_p.random,
                        CH_p.legacy_session_id, CH_p.cipher_suite,
                        CH_p.legacy_compression_methods, CH_p.extensions))
end process

-------------------------------------------------------------------------------

process EncryptedExtensions_TP [encryptede: EE] is
  encryptede (EncryptedExtensions ({}))
end process

-------------------------------------------------------------------------------

process CertificateRequest_TP [certificater: CR] is
  var v_certificate_context: CertificateRequestContext,
      v_extensions: Extensions in
    v_certificate_context := {};
    v_extensions := {Extension (void_filters,
                                OIDFilterExtension ({peer ({x00}, {x00})}))};
    v_extensions :=
      add_extension (Extension (signature_algorithms,
                                SignatureSchemeList ({rsa_pkcs1_sha256})),
                     v_extensions);
    certificater (CertificateRequest (v_certificate_context, v_extensions))
  end var
end process

-------------------------------------------------------------------------------

process Certificate_S_TP [certificate: C]
                      (in out C_p: Certificate,
                       CH_p: ClientHello, CR_p: CertificateRequest) is
  var ed: ExtensionData in
    C_p := C_p.{crc => CR_p.certificate_context};
    -- X509
    ed := CertificateType (x509);
    C_p := C_p.{certificate_list =>
                  cons (add_Entry (ed.ct, CH_p), C_p.certificate_list)};
    certificate (Certificate(C_p.crc, C_p.certificate_list))
  end var
end process

-------------------------------------------------------------------------------

process Certificate_C_TP [certificate: C]
                      (in out C_p: Certificate, CH_p: ClientHello,
                      CR_p: CertificateRequest) is
  use CH_p; use CR_p;
  C_p := C_p.{certificate_list => {}, crc => {}};
  certificate (Certificate(C_p.crc, C_p.certificate_list))
end process

end module
\end{lstlisting}
\label{ap:tp}

\clearpage
\subsection{Test purpose I}
\label{ap:tp1}
\begin{lstlisting}[language=LNT]
module handshake_TP (common_TP_interactions) is

-------------------------------------------------------------------------------

process Client [clientHello_c: CH, serverHello_c: SH,
                encryptedExtensions_c: EE, certificateRequest_c: CR,
                certificate_c_c, certificate_s_c: C,
                certificateVerify_s_c: CV, finished_c_c,
                finished_s_c: F] is
  var alert: AlertType, CH_p: CLientHello, C_C_p: Certificate,
      CR_p: CertificateRequest, HRR_P: HelloRetryRequest,
      is_helloRequest: BOOL in
    -- Initialisation of variables
    CH_p := ClientHello (T_NULL, undefined, T_NULL, {}, T_NULL, {});
    HRR_P := HelloRetryRequest (T_NULL, {}, {});
    CR_p := CertificateRequest ({},{});
    use CR_p;
    C_C_p := Certificate ({},{});
    is_helloRequest := false;
    -- client key exchange
    ClientHello_TP [clientHello_c] (is_helloRequest, !?CH_p, HRR_P, ?alert);
    use alert;
    serverHello_c (?any ServerHello);
    encryptedExtensions_c (?any EncryptedExtensions);
    -- protocol messages sent in the right and classical order
    certificateRequest_c (?CR_p);
    -- E. WAIT_Certificate
    certificate_s_c (?any Certificate);
    certificateVerify_s_c (?any CertificateVerify);
    finished_s_c (?any Finished);
    -- client authentification
    Certificate_C_TP [certificate_c_c] (!?C_C_p, CH_p, CR_p);
    use C_C_p;
    Finished_C [finished_c_c]
  end var
end process

-------------------------------------------------------------------------------

process Server [clientHello_c: CH, serverHello_c: SH,
                encryptedExtensions_c: EE,
                certificateRequest_c: CR, certificate_c_c, certificate_s_c: C,
                certificateVerify_s_c: CV,
                finished_c_c, finished_s_c: F] is
  var alert: AlertType, CH_p: CLientHello, C_S_p: Certificate,
      CR_p: CertificateRequest in
    -- Initialisation of variables
    alert := undefined; use alert;
    CR_p := CertificateRequest ({},{});
    C_S_p := Certificate ({},{});
    -- client key exchange
    clientHello_c (?CH_p);
    ServerHello [serverHello_c];
    EncryptedExtensions_TP [encryptedExtensions_c];
    -- protocol messages sent in the right and classical order
    CertificateRequest_TP [certificateRequest_c];
    Certificate_S_TP [certificate_s_c](!?C_S_p, CH_p, CR_p);
    use C_S_p;
    CertificateVerify_S [certificateVerify_s_c];
    Finished_S [finished_s_c];
    -- client authentification
    certificate_c_c (?any Certificate);
    finished_c_c (?any Finished)
  end var
end process

-------------------------------------------------------------------------------

process MAIN [clientHello_c: CH, serverHello_c: SH,
              encryptedExtensions_c: EE, certificateRequest_c: CR,
              certificate_c_c, certificate_s_c: C,
              certificateVerify_s_c: CV, finished_c_c, finished_s_c: F,
              TESTOR_ACCEPT: none] is
  par
    clientHello_c, serverHello_c, encryptedExtensions_c,
    certificateRequest_c, certificate_c_c,
    certificate_s_c, certificateVerify_s_c,finished_c_c, finished_s_c ->
      Client [clientHello_c, serverHello_c, encryptedExtensions_c,
              certificateRequest_c, certificate_c_c, certificate_s_c,
              certificateVerify_s_c, finished_c_c, finished_s_c]
  ||
    clientHello_c, serverHello_c, encryptedExtensions_c,
    certificateRequest_c, certificate_c_c, certificate_s_c,
    certificateVerify_s_c, finished_c_c, finished_s_c ->
      Server [clientHello_c, serverHello_c, encryptedExtensions_c,
              certificateRequest_c, certificate_c_c, certificate_s_c,
              certificateVerify_s_c, finished_c_c, finished_s_c]
  end par;
  TESTOR_ACCEPT
end process

end module
\end{lstlisting}

\clearpage
\subsection{Test purpose II}
\label{ap:tp2}
\begin{lstlisting}[language=LNT]
module handshake_helloRequest_TP (common_TP_interactions) is

-------------------------------------------------------------------------------

process Client [clientHello_c: CH, serverHello_c: SH,
                helloRetryRequest_c: HRR, encryptedExtensions_c: EE,
                certificateRequest_c: CR, certificate_c_c, certificate_s_c: C,
                certificateVerify_s_c: CV,
                finished_c_c, finished_s_c: F] is
  var alert: AlertType, CH_p: CLientHello, C_C_p: Certificate,
      CR_p: CertificateRequest, HRR_P: HelloRetryRequest,
      is_helloRequest: BOOL in
    -- Initialisation of variables
    CH_p := ClientHello (T_NULL, undefined, T_NULL, {}, T_NULL, {});
    HRR_P := HelloRetryRequest (T_NULL, {}, {});
    C_C_p := Certificate ({},{});
    is_helloRequest := false;
    -- client key exchange
    ClientHello_TP [clientHello_c] (is_helloRequest, !?CH_p, HRR_P, ?alert);
    use alert;
    -- protocol messages sent in the right order with incorrect Key shared
    helloRetryRequest_c (?HRR_P);
    is_helloRequest := true;
    -- client key exchange
    ClientHello_TP [clientHello_c] (is_helloRequest, !?CH_p, HRR_P, ?alert);
    use alert;
    serverHello_c (?any ServerHello);
    encryptedExtensions_c (?any EncryptedExtensions);
    -- protocol messages sent in the right and classical order
    certificateRequest_c (?CR_p);
    certificate_s_c (?any Certificate);
    certificateVerify_s_c (?any CertificateVerify);
    finished_s_c (?any Finished);
    -- client authentification
    Certificate_C_TP [certificate_c_c] (!?C_C_p, CH_p, CR_p);
    use C_C_p;
    Finished_C [finished_c_c]
  end var
end process

-------------------------------------------------------------------------------

process Server [clientHello_c: CH, serverHello_c: SH,
                helloRetryRequest_c: HRR, encryptedExtensions_c: EE,
                certificateRequest_c: CR, certificate_c_c,
                certificate_s_c: C, certificateVerify_s_c: CV,
                finished_c_c, finished_s_c: F] is
  var CH_p: CLientHello, C_S_p: Certificate, CR_p: CertificateRequest,
      HRR_P: HelloRetryRequest in
    -- Initialisation of variables
    HRR_P := HelloRetryRequest (T_NULL, {}, {});
    CR_p := CertificateRequest ({},{});
    C_S_p := Certificate ({},{});
    -- client key exchange
    clientHello_c (?CH_p);
    use CH_P;
    -- protocol messages sent in the right order with incorrect Key shared
    HelloRetryRequest [helloRetryRequest_c](!?HRR_P);
    use HRR_P;
    -- client key exchange
    clientHello_c (?CH_p);
    ServerHello [serverHello_c];
    EncryptedExtensions_TP [encryptedExtensions_c];
    -- protocol messages sent in the right and classical order
    CertificateRequest_TP [certificateRequest_c];
    Certificate_S_TP [certificate_s_c](!?C_S_p, CH_p, CR_p);
    use C_S_p;
    CertificateVerify_S [certificateVerify_s_c];
    Finished_S [finished_s_c];
    -- client authentification
    certificate_c_c (?any Certificate);
    finished_c_c (?any Finished)
  end var
end process

-------------------------------------------------------------------------------

process MAIN [clientHello_c: CH, serverHello_c: SH, helloRetryRequest_c: HRR,
              encryptedExtensions_c: EE, certificateRequest_c: CR,
              certificate_s_c, certificate_c_c: C,
              certificateVerify_s_c: CV, finished_c_c, finished_s_c: F,
              TESTOR_ACCEPT: none] is
  par
    clientHello_c, serverHello_c, helloRetryRequest_c,
    encryptedExtensions_c, certificateRequest_c, certificate_c_c,
    certificate_s_c, certificateVerify_s_c,
    finished_c_c, finished_s_c ->
      Client [clientHello_c, serverHello_c, helloRetryRequest_c,
              encryptedExtensions_c, certificateRequest_c, certificate_c_c,
              certificate_s_c, certificateVerify_s_c,
              finished_c_c, finished_s_c]
  ||
    clientHello_c, serverHello_c, helloRetryRequest_c,
    encryptedExtensions_c, certificateRequest_c, certificate_c_c,
    certificate_s_c, certificateVerify_s_c,
    finished_c_c, finished_s_c ->
      Server [clientHello_c, serverHello_c, helloRetryRequest_c,
              encryptedExtensions_c, certificateRequest_c, certificate_c_c,
              certificate_s_c, certificateVerify_s_c,
              finished_c_c, finished_s_c]
  end par;
  TESTOR_ACCEPT
end process

end module
\end{lstlisting}

\clearpage
\subsection{Test purpose III}
\label{ap:tp3}
\begin{lstlisting}[language=LNT]
module handshake_alert_TP (common_TP_interactions) is

-------------------------------------------------------------------------------

process Client [clientHello_c: CH, serverHello_c: SH,
                encryptedExtensions_c: EE, certificateRequest_c: CR,
                certificate_s_c: C] is
  var alert: AlertType, CH_p: CLientHello, HRR_P: HelloRetryRequest,
      is_helloRequest : BOOL in
    -- Initialisation of variables
    CH_p := ClientHello (T_NULL, undefined, T_NULL, {}, T_NULL, {});
    HRR_P := HelloRetryRequest (T_NULL, {}, {});
    is_helloRequest := false;
    -- client key exchange
    ClientHello_TP [clientHello_c] (is_helloRequest, !?CH_p, HRR_P, ?alert);
    use alert;
    serverHello_c (?any ServerHello);
    encryptedExtensions_c (?any EncryptedExtensions);
    -- Recv CertificateRequest
    certificateRequest_c (?any CertificateRequest);
    -- E. WAIT_Certificate
    certificate_s_c (?any Certificate);
    ClientHello_TP [clientHello_c] (is_helloRequest, !?CH_p, HRR_P, ?alert);
    use CH_p; use alert
  end var
end process

-------------------------------------------------------------------------------

process Server [clientHello_c: CH, serverHello_c: SH,
                encryptedExtensions_c: EE, certificateRequest_c: CR,
                certificate_s_c: C, alert_c: A] is
  var alert: AlertType, CH_p: CLientHello, C_S_p: Certificate,
      CR_p: CertificateRequest in
    -- Initialisation of variables
    alert := undefined;
    use alert;
    CR_p := CertificateRequest ({},{});
    C_S_p := Certificate ({},{});
    -- client key exchange
    -- Rec ClientHello
    -- B. RECVD_CH
    clientHello_c (?CH_p);
    ServerHello [serverHello_c];
    EncryptedExtensions_TP [encryptedExtensions_c];
    CertificateRequest_TP [certificateRequest_c];
    Certificate_S_TP [certificate_s_c](!?C_S_p, CH_p, CR_p);
    use C_S_p;
    clientHello_c (?any CLientHello);
    alert := unexpected_message;
    -- abort the handshake with an "unexpected_message" alert
    alert_c (alert)
  end var
end process

-------------------------------------------------------------------------------

process MAIN [clientHello_c: CH, serverHello_c: SH,
              encryptedExtensions_c: EE, certificateRequest_c: CR,
              certificate_s_c: C, alert_c: A, TESTOR_ACCEPT: none] is
  par
    clientHello_c, serverHello_c, encryptedExtensions_c,
    certificateRequest_c, certificate_s_c ->
      Client [clientHello_c, serverHello_c, encryptedExtensions_c,
              certificateRequest_c, certificate_s_c]
  ||
    clientHello_c, serverHello_c, encryptedExtensions_c,
    certificateRequest_c, certificate_s_c ->
      Server [clientHello_c, serverHello_c, encryptedExtensions_c,
              certificateRequest_c, certificate_s_c, alert_c]
  end par;
  TESTOR_ACCEPT
end process

end module
\end{lstlisting}

\end{document}